\documentclass[english]{article}
\usepackage[T1]{fontenc}
\usepackage{graphicx}
\usepackage[latin9]{inputenc}
\usepackage{amsmath,mathrsfs,mathtools,amsthm}
\usepackage{amsmath,amsfonts,amssymb,epsfig,multicol,multirow}
\usepackage{rotating,hhline,stmaryrd}
\usepackage{float}
\usepackage{array}
\newcommand{\Proj}{\text{Proj}}
\newtheorem{definition}{Definition}
\newtheorem{theorem}{Theorem}
\newtheorem{proposition}{Proposition}

\newcommand{\ob}{\mbox{$\overline{\omega}$}}
\newcommand{\om}{\mbox{$\omega$}}

\newcommand{\C}{\mbox{$\cal C$}}

\newcommand{\ben}{\begin{equation*}}
\newcommand{\een}{\end{equation*}}

\begin{document}

\title{Projection decoding of some binary optimal linear codes of lengths 36 and 40}

\author{Lucky Galvez \thanks{ Department of Mathematics,
Sogang University,
Seoul 04107, South Korea; and Institute
of Mathematics, University of the Philippines Diliman, Quezon City 1101, Philippines
{Email: \tt  le$\_$galvez@yahoo.com.ph}}
\and Jon-Lark Kim\thanks{corresponding author, Department of Mathematics,
Sogang University,
Seoul 04107, South Korea.
{Email: \tt jlkim@sogang.ac.kr},
J.-L. Kim was supported by Basic Research Program through the National Research Foundation of Korea (NRF) funded by the Ministry of Education (NRF-2016R1D1A1B03933259).}
}

\date{}

\maketitle

\begin{abstract}
Practically good error-correcting codes should have good parameters and efficient decoding algorithms. Some algebraically defined good codes such as cyclic codes, Reed-Solomon codes, and Reed-Muller codes have nice decoding algorithms. However, many optimal linear codes do not have an efficient decoding algorithm except for the general syndrome decoding which requires a lot of memory. Therefore, it is a natural question whether which optimal linear codes have an efficient decoding. We show that two binary optimal $[36,19,8]$ linear codes and two binary optimal $[40,22,8]$ codes have an efficient decoding algorithm. There was no known efficient decoding algorithm for the binary optimal $[36,19,8]$ and $[40,22,8]$ codes. We project them onto the much shorter length linear $[9,5,4]$ and $[10, 6, 4]$ codes over $GF(4)$, respectively. This decoding algorithms, called {\em projection decoding}, can correct errors of weight up to 3. These $[36,19,8]$ and $[40,22,8]$ codes respectively have more codewords than any optimal self-dual $[36, 18, 8]$ and $[40,20,8]$ codes for given length and minimum weight, implying that these codes more practical.
\end{abstract}

\noindent
{\bf keywords} codes, optimal codes, self-dual codes, projection decoding

\section{Introduction}

Coding theory or the theory of error-correcting codes requires a lot of mathematical concepts but has wide applications to data storage, satellite communication, smart phone, and High Definition TV. The well known classes of (linear) codes include Reed-Solomon codes, Reed-Muller codes, turbo codes, LDPC codes, Polar codes, network codes, quantum codes, and DNA codes. One of the most important reasons why these codes are useful is that they have fast and efficient decoding algorithms.

\medskip

A {\em linear $[n,k]$ code} over $GF(q)$ or $\mathbb F_q$ is a $k$-dimensional subspace of $\mathbb F_q^n$. The {\em dual} of $\C$ is $\C^{\perp}=\{x \in \mathbb F_q^n~|~ x \cdot c = 0 {\mbox{ for any }} c \in \C\}$, where the dot product is either a usual inner product or a Hermitian inner product. A linear code $\C$ is called {\em self-dual} if $\C=\C^{\perp}$. If $q=2$, then $C$ is called {\em binary}. A binary self-dual code is called {\em doubly-even} if all codewords
have weight $\equiv 0 \pmod 4$ and {\em singly-even} if some codeword
has a weight $\equiv 2 \pmod 4$.
If $q=4$, let $GF(4)=\{0, 1, \om, \ob\}$, where $\ob=\om^2 = \om +1$. It is more natural to consider the Hermitian inner product $\left <, \right>$ on $GF(4)^n$:
for $x=(x_1, x_2, \dots, x_n)$ and $y=(y_1, y_2, \dots, y_n)$ in $GF(4)^n$, $\left <x,y \right>= \sum_{i=1}^n x_i \overline{y_i}$, where $\overline{a}= a^2$.

\medskip

Researchers have tried to find an efficient decoding algorithms for some optimal linear codes of lengths up to 40. For example,
Pless~\cite{Pless86} showed an efficient decoding of the $[24,12,8]$ extended binary self-dual Golay code by projecting it onto the $[6, 3, 4]$ hexacode over $GF(4)$.  Later, Gaborit, Kim and Pless~\cite{GKP03,KimPless02} showed that a similar projection can be done for some singly even and doubly-even self-dual binary $[32, 16, 8]$ codes including the binary Reed Muller code.
Recently, Kim and Lee \cite{KimLee16} gave two algorithms for the projection decoding of a binary extremal self-dual code of length 40. The idea was to use the projection of the said code onto a Hermitian self-dual code over $GF(4)$. One of the two algorithms called syndrome decoding uses the syndrome decoding of the shorter code over $GF(4)$. Most examples for the projection decoding were extremal self-dual codes. Therefore it is natural to ask whether such a projection decoding can be done for a non-self-dual code which has larger dimension than a self-dual code with the same length and minimum distance. In this paper, we show that this is possible.

\medskip

The purpose of this paper is to show how to decode efficiently a binary optimal $[36,19,8]$ linear code and a binary optimal $[40,22,8]$ code by projecting them onto the much shorter length linear $[9,5,4]$ and $[10, 6, 4]$ codes over $GF(4)$, respectively. This decoding algorithm, which we will call \emph{projection decoding} can correct errors of weight up to $\lfloor \frac{(8-1)}{2}\rfloor =3$. It can be seen that the decoding algorithm presented in this paper is a generalization of the syndrome projection decoding in \cite{KimLee16} since this algorithm can be used to decode any linear code with a projection into an additive or a linear code over $GF(4)$ for errors of weight at most $3$. Our decoding is the first time to decode those two optimal $[36, 19, 8]$ and $[40, 22, 8]$ codes, whose parameters are better than any optimal self-dual $[36, 18, 8]$ and $[40, 20, 8]$ codes.

\medskip

This paper is organized as follows. In Section 2, we introduce a way to project a binary code of length $4n$ into an additive code of length $n$ over $GF(4)$. We also mention some properties of this projection as given in \cite{AmraniBeery01, KMP03}. In Section 3, we show explicitly how to construct  $[36,19,8]$ and $[40,22,8]$ binary optimal codes having projections onto additive codes over $GF(4)$. Using this projection, we give a handy decoding algorithm to decode the said binary optimal codes in Section \ref{sec:decoding}. This decoding algorithm exploits the properties of codes with projection onto an additive code over $GF(4)$ to locate the errors in the noisy codewords, assuming not more than 3 errors occurred. This algorithm has low complexity and can be done with hand.
In Section 5, we conclude by showing working examples of how the decoding is carried out.

\section{Projection of binary linear codes} \label{sec:projection}

Let $v\in GF(2)^{4m}$. We associate to $v$ a rectangular array $\mathbf{v}$ of zeros and ones. If we label the rows of the array with the elements of $GF(4)$, then the inner product of a column  of the array with the row labels is an element of $GF(4)$. Thus, we obtain a corresponding element of $GF(4)^m$, which we call the \emph{projection} of $v$, denoted $\Proj(\mathbf{v})$. We show this by the following
example. Let $v=(1000~\,0100~\,0010~\,0001~\,0110~\,1110~\,0101~\,0111~\,1111) \in GF(2)^{36}$.
Then we write $v$ column-wisely as follows.
\[
\mathbf{v}=\begin{array}{c|ccccccccc}
 & 1 & 2 & 3 & 4 & 5 & 6 & 7 & 8 & 9\\
 \hline
0 & 1 & 0 & 0 & 0 & 0 & 1 & 0 & 0 & 1\\
1 & 0 & 1 & 0 & 0 & 1 & 1 & 1 & 1 & 1\\
\omega & 0 & 0 & 1 & 0 & 1 & 1 & 0 & 1 & 1\\
\overline{\omega}  & 0 & 0 & 0 & 1 & 0 & 0 & 1 & 1 & 1  \\ \hline
   & \mathbf{0} & \mathbf{1} & \mathbf{\mathbf{\omega}} & \mathbf{\overline{\omega}} & \mathbf{\mathbf{\overline{\omega}}} & \mathbf{\mathbf{\overline{\omega}}} & \mathbf{\mathbf{\omega}} & \mathbf{0} & \mathbf{0}
\end{array}
\]
We define the parity of the column to be {\em even} (or {\em odd}) if an even (or odd) number of ones exits in the column, and the parity of the first row is defined similarly.

We call $C_4 \subset GF(4)^n$ an {\em additive code} of length $n$ over $GF(4)$ if it is closed under addition. Therefore, a linear code over $GF(4)$ is automatically additive, but not the converse.

\begin{definition} \label{def:projectionO}
{\em Let $S$ be a subset of $GF(2)^{4m}$ and $C_4$ an additive code over $GF(4)$ of length $m$. Then $S$ is said to have a {\em projection $O$ onto $C_4$} if for any $v \in S$,
\begin{enumerate}
\item[(i)] $\Proj(\mathbf{v}) \in C_4$.
\item[(ii)] the columns of $\mathbf{v}$ are of the same parity, i.e., the columns are either all even or all odd
\item[(iii)] the parity of the first row of $\mathbf{v}$ is the same as the parity of the columns
\end{enumerate}
}

\end{definition}

The main advantage of this projection is that we can decode the long binary code by decoding its projection, thus generally decreasing the complexity of the decoding process. Several authors have exhibited this fact.

Another similar projection is defined in \cite{KMP03}, called projection $E$.

\begin{definition} \label{def:projectionE}
{\em Using the same notation as in Definition \ref{def:projectionO}, $S$ is said to have a {\em projection $E$ onto $C_4$} if conditions $(i)$ and $(ii)$ are satisfied together with the additional condition:
\begin{enumerate}
\item[(iii)'] the parity of the first row of $\mathbf{v}$ is always even.
\end{enumerate}
}
\end{definition}

Let $C_4$ be an additive code of length $m$ over $GF(4)$. Consider the map $\phi: GF(4) \to GF(2)^4$ such that $\phi(0) = 0000$, $\phi(1) = 0011$, $\phi(\omega) = 0101$, and $\phi(\overline{\omega}) = 0110$. Define $$\phi(C_4) = \{(\phi(c_1),\phi(c_2),\ldots,\phi(c_m)): (c_1,c_2, \ldots, c_m) \in C_4 \}.$$
Let $d$ be the code consisting of all even sums of weight $4$ vectors whose ones appear in the same column in the projection, together with one additional vector $d_1 = (1000\,1000\,\ldots\,1000)$ if $m$ is odd and $d_2 = (1000\,1000\, \ldots \, 0111)$ if $m$ is even.
In \cite{KMP03}, the following constructions were given:
\begin{enumerate}
\item[] {\bf Construction $O$}: $\rho_O(C_4) = \phi(C_4) + d$

\item[] {\bf Construction $E$}: $\rho_E(C_4) = \phi(C_4) + d^\prime$
\end{enumerate}

\noindent where $d'$ is the code consisting of $d_1$ if $m$ is even and $d_2$ if $m$ is odd.

The following result also was also given in \cite{KMP03}.
\begin{theorem}\label{thm:proj}
Let $C_4$ be an additive $(m,2^r)$ code over $GF(4)$. Then
\begin{enumerate}
\item $\rho_O(C_4)$ and $\rho_E(C_4)$ are binary linear $[4m,m+r]$ code having projection $O$ and $E$, respectively, onto $C_4$.
\item any binary linear code having projection $O$ or $E$ onto $C_4$ can be constructed in this way.
\end{enumerate}
\end{theorem}

\section{Construction of binary optimal $[36,19,8]$ and $[40,22,8]$ codes}\label{sec:construction}

In this section, we apply the constructions given in the previous section to construct binary optimal codes of lengths $36$ and  $40$. We were able to obtain two inequivalent codes for each length.

Let $C_{4}^9$ be a $(9,2^{10})$ additive code over $GF(4)$ with the following generator matrix 
{\small \[ G(C_4^9) = \left[ \begin{array}{ccccccccc}
1 &  0 &  0 &  0 &  0 & \overline{\omega}  & 1 &  1 &  1 \\
  \omega &   0 &  0 &  0 &  0 &  1 &  \omega &  \omega &  \omega \\
 0  & 1 &  0  & 0 &  0  & 1 &  \omega & \overline{\omega}  & 0 \\
 0 &  \omega &   0 &  0 &  0 & \omega & \overline{\omega}  & 1 &  0 \\
  0 &  0 &  1 &  0 &  0 &  0  & 1 &  \omega & \overline{\omega} \\
 0 &  0 &  \omega &  0 &  0 &  0 &  \omega & \overline{\omega} & 1 \\
0 &  0 &  0 &  1 &  0 & \overline{\omega} & \overline{\omega}  & 0 &  1 \\
 0 &  0 &  0 &  \omega & 0 &  1 &  1 &  0 &  \omega \\
 0 &  0 &  0 &  0 &  1 &  1 &  \omega &  1 &  1 \\
  0 &  0 &  0 &  0 &  \omega &   \omega & \overline{\omega} &  \omega &  \omega
  \end{array} \right]. \]}

In fact the rows consisting of the odd indexed rows of $G(C_4^9)$ form a $[9,5,4]$-linear code over $GF(4)$ which can be found in MAGMA with the name of $BKLC(GF(4),9,5)$. The code $C_4^9$ has weight distribution $A_4 = 51, A_5 = 135, A_6=210, A_7=318, A_8=234, A_9 = 75$.

\medskip

Let $C_4^{10}$ be the $(10,2^{12})$ additive code over $GF(4)$ generated by the following matrix
{\small \[ G(C_4^{10}) = \left[ \begin{array}{cccccccccc}
 1  & 0  & 0  & 0  & 0 & 0 &  \omega &  0 &  \omega & \overline{\omega} \\
  \omega  & 0  & 0  & 0  & 0 &  0 & \overline{\omega} &  0 & \overline{\omega} &  1 \\
  0  & 1  & 0   &0  & 0 &  0 & \overline{\omega} &  1 &  1 &  1 \\
  0  & w  & 0  & 0 &  0 &  0 &  1 &  \omega &  \omega &  \omega \\
  0  & 0  & 1  & 0  & 0 &  0 &  1 &  \omega & \overline{\omega} &  0 \\
  0  & 0  & \omega  & 0 &  0 &  0 &  \omega & \overline{\omega} &  1 &  0 \\
  0  & 0  & 0  & 1  & 0 &  0 &  0 &  1 &  \omega & \overline{\omega} \\
  0  & 0  & 0  & \omega  & 0 &  0 &  0 &  \omega & \overline{\omega} &  1 \\
  0  & 0  & 0  & 0  & 1  & 0 & \overline{\omega} & \overline{\omega} &  0 &  1 \\
  0  & 0  & 0  & 0 &  \omega &  0 &  1 &  1 &  0 &  \omega \\
  0  & 0  & 0  & 0  & 0  & 1 &  1 &  \omega  & 1 &  1 \\
  0  & 0  & 0  & 0  & 0 &  \omega &  \omega & \overline{\omega} &  \omega &  \omega
  \end{array} \right]. \]}
The minimum distance of both of these codes is $4$.

The rows consisting of the odd indexed rows of $G(C_4^{10})$ form a $[10,6,4]$-linear code over $GF(4)$ which can be found in MAGMA with the name of $BKLC(GF(4),10,6)$. The code $C_4^9$ has weight distribution $A_4 = 87, A_5 = 258, A_6=555, A_7=1020, A_8=1200, A_9 = 738, A_{10}=237$.

\medskip

Denote by  $\mathcal{C}_O^{36}$ and $\mathcal{C}_O^{40}$ the binary linear codes obtained from the additive codes $C_4^9$ and $C_4^{10}$, respectively,  over $GF(4)$ by construction $O$. That is, $\mathcal{C}_O^{36}=\rho_{O}(C_{4}^9)$ and $\mathcal{C}_O^{40} = \rho_O(C_4^{10})$. Their generator matrices are given below. These two codes constructed are inequivalent and both have projection $O$ on to $C_4^9$ and $C_4^{10}$, respectively.

{\small \[ G(\mathcal{C}_O^{36}) = \left[ \begin{array} {c}
1 0 0 0 0 0 0 1 0 0 0 1 0 0 0 1 0 0 0 1 0 0 1 0 0 1 0 0 0 1 1 1 0 0 1 0 \\
0 1 0 0 0 0 0 1 0 0 0 1 0 0 0 1 0 0 0 1 0 1 0 0 0 1 1 1 0 1 0 0 1 1 1 0 \\
0 0 1 0 0 0 0 1 0 0 0 1 0 0 0 1 0 0 0 1 0 0 0 1 0 0 0 1 0 0 1 0 1 0 0 0 \\
0 0 0 1 0 0 0 1 0 0 0 1 0 0 0 1 0 0 0 1 0 1 1 1 0 0 1 0 0 0 0 1 1 0 1 1 \\
0 0 0 0 1 0 0 1 0 0 0 0 0 0 0 0 0 0 0 0 0 1 1 0 0 0 1 1 0 1 0 1 1 1 1 1 \\
0 0 0 0 0 1 0 1 0 0 0 0 0 0 0 0 0 0 0 0 0 1 0 1 0 1 1 0 0 0 1 1 0 0 0 0 \\
0 0 0 0 0 0 1 1 0 0 0 0 0 0 0 0 0 0 0 0 0 0 1 1 0 1 0 1 0 1 1 0 0 0 0 0 \\
0 0 0 0 0 0 0 0 1 0 0 1 0 0 0 0 0 0 0 0 0 0 0 0 0 1 1 0 0 0 1 1 1 0 1 0 \\
0 0 0 0 0 0 0 0 0 1 0 1 0 0 0 0 0 0 0 0 0 0 0 0 0 1 0 1 0 1 1 0 0 0 1 1 \\
0 0 0 0 0 0 0 0 0 0 1 1 0 0 0 0 0 0 0 0 0 0 0 0 0 0 1 1 0 1 0 1 0 1 1 0 \\
0 0 0 0 0 0 0 0 0 0 0 0 1 0 0 1 0 0 0 0 0 1 0 1 0 1 0 1 0 0 0 0 1 0 0 1 \\
0 0 0 0 0 0 0 0 0 0 0 0 0 1 0 1 0 0 0 0 0 0 1 1 0 0 1 1 0 0 0 0 0 1 0 1 \\
0 0 0 0 0 0 0 0 0 0 0 0 0 0 1 1 0 0 0 0 0 1 1 0 0 1 1 0 0 0 0 0 0 0 1 1 \\
0 0 0 0 0 0 0 0 0 0 0 0 0 0 0 0 1 0 0 1 0 1 1 0 0 0 1 1 0 1 1 0 1 0 0 1 \\
0 0 0 0 0 0 0 0 0 0 0 0 0 0 0 0 0 1 0 1 0 1 0 1 0 1 1 0 0 1 0 1 0 1 0 1 \\
0 0 0 0 0 0 0 0 0 0 0 0 0 0 0 0 0 0 1 1 0 0 1 1 0 1 0 1 0 0 1 1 0 0 1 1 \\
0 0 0 0 0 0 0 0 0 0 0 0 0 0 0 0 0 0 0 0 1 1 1 1 0 0 0 0 0 0 0 0 1 1 1 1 \\
0 0 0 0 0 0 0 0 0 0 0 0 0 0 0 0 0 0 0 0 0 0 0 0 1 1 1 1 0 0 0 0 1 1 1 1 \\
0 0 0 0 0 0 0 0 0 0 0 0 0 0 0 0 0 0 0 0 0 0 0 0 0 0 0 0 1 1 1 1 1 1 1 1
\end{array} \right] \]}

{\small \[ G(\mathcal{C}_O^{40}) = \left[ \begin{array} {c}
1 0 0 0 0 0 0 1 0 0 0 1 0 0 0 1 0 0 0 1 0 0 0 1 0 1 1 1 0 0 1 0 0 0 0 1 0 1 0 0 \\
0 1 0 0 0 0 0 1 0 0 0 1 0 0 0 1 0 0 0 1 0 0 0 1 0 0 1 0 0 0 1 0 0 1 0 0 1 1 0 1 \\
0 0 1 0 0 0 0 1 0 0 0 1 0 0 0 1 0 0 0 1 0 0 0 1 0 0 0 1 0 0 1 0 0 1 1 1 1 0 0 0 \\
0 0 0 1 0 0 0 1 0 0 0 1 0 0 0 1 0 0 0 1 0 0 0 1 0 1 0 0 0 0 1 0 0 0 1 0 1 1 1 0 \\
0 0 0 0 1 0 0 1 0 0 0 0 0 0 0 0 0 0 0 0 0 0 0 0 0 1 0 1 0 1 1 0 0 1 1 0 1 0 0 1 \\
0 0 0 0 0 1 0 1 0 0 0 0 0 0 0 0 0 0 0 0 0 0 0 0 0 0 1 1 0 1 0 1 0 1 0 1 0 1 0 1 \\
0 0 0 0 0 0 1 1 0 0 0 0 0 0 0 0 0 0 0 0 0 0 0 0 0 1 1 0 0 0 1 1 0 0 1 1 0 0 1 1 \\
0 0 0 0 0 0 0 0 1 0 0 1 0 0 0 0 0 0 0 0 0 0 0 0 0 1 1 0 0 0 1 1 0 1 0 1 1 1 1 1 \\
0 0 0 0 0 0 0 0 0 1 0 1 0 0 0 0 0 0 0 0 0 0 0 0 0 1 0 1 0 1 1 0 0 0 1 1 0 0 0 0 \\
0 0 0 0 0 0 0 0 0 0 1 1 0 0 0 0 0 0 0 0 0 0 0 0 0 0 1 1 0 1 0 1 0 1 1 0 0 0 0 0 \\
0 0 0 0 0 0 0 0 0 0 0 0 1 0 0 1 0 0 0 0 0 0 0 0 0 0 0 0 0 1 1 0 0 0 1 1 1 0 1 0 \\
0 0 0 0 0 0 0 0 0 0 0 0 0 1 0 1 0 0 0 0 0 0 0 0 0 0 0 0 0 1 0 1 0 1 1 0 0 0 1 1 \\
0 0 0 0 0 0 0 0 0 0 0 0 0 0 1 1 0 0 0 0 0 0 0 0 0 0 0 0 0 0 1 1 0 1 0 1 0 1 1 0 \\
0 0 0 0 0 0 0 0 0 0 0 0 0 0 0 0 1 0 0 1 0 0 0 0 0 1 0 1 0 1 0 1 0 0 0 0 1 0 0 1 \\
0 0 0 0 0 0 0 0 0 0 0 0 0 0 0 0 0 1 0 1 0 0 0 0 0 0 1 1 0 0 1 1 0 0 0 0 0 1 0 1 \\
0 0 0 0 0 0 0 0 0 0 0 0 0 0 0 0 0 0 1 1 0 0 0 0 0 1 1 0 0 1 1 0 0 0 0 0 0 0 1 1 \\
0 0 0 0 0 0 0 0 0 0 0 0 0 0 0 0 0 0 0 0 1 0 0 1 0 1 1 0 0 0 1 1 0 1 1 0 1 0 0 1 \\
0 0 0 0 0 0 0 0 0 0 0 0 0 0 0 0 0 0 0 0 0 1 0 1 0 1 0 1 0 1 1 0 0 1 0 1 0 1 0 1 \\
0 0 0 0 0 0 0 0 0 0 0 0 0 0 0 0 0 0 0 0 0 0 1 1 0 0 1 1 0 1 0 1 0 0 1 1 0 0 1 1 \\
0 0 0 0 0 0 0 0 0 0 0 0 0 0 0 0 0 0 0 0 0 0 0 0 1 1 1 1 0 0 0 0 0 0 0 0 1 1 1 1 \\
0 0 0 0 0 0 0 0 0 0 0 0 0 0 0 0 0 0 0 0 0 0 0 0 0 0 0 0 1 1 1 1 0 0 0 0 1 1 1 1 \\
0 0 0 0 0 0 0 0 0 0 0 0 0 0 0 0 0 0 0 0 0 0 0 0 0 0 0 0 0 0 0 0 1 1 1 1 1 1 1 1
\end{array} \right] \]}

Similarly, we apply construction $E$ on the codes $C_4^9$ and $C_4^{10}$. We obtain two inequivalent codes $\mathcal{C}_E^{36} = \rho_E(C_4^9)$ and $\mathcal{C}_E^{40} = \rho_E(C_4^{10})$ with projection $E$ on to $C_4^9$ and $C_4^{10}$, respectively. Their generator matrices are as follows.
{\small \[ G(\mathcal{C}_E^{36}) = \left[ \begin{array} {c}
1 0 0 0 0 0 0 1 0 0 0 1 0 0 0 1 0 0 0 1 0 0 1 0 0 1 0 0 0 1 1 1 1 1 0 1 \\
0 1 0 0 0 0 0 1 0 0 0 1 0 0 0 1 0 0 0 1 0 1 0 0 0 1 1 1 0 1 0 0 0 0 0 1 \\
0 0 1 0 0 0 0 1 0 0 0 1 0 0 0 1 0 0 0 1 0 0 0 1 0 0 0 1 0 0 1 0 0 1 1 1 \\
0 0 0 1 0 0 0 1 0 0 0 1 0 0 0 1 0 0 0 1 0 1 1 1 0 0 1 0 0 0 0 1 0 1 0 0 \\
0 0 0 0 1 0 0 1 0 0 0 0 0 0 0 0 0 0 0 0 0 1 1 0 0 0 1 1 0 1 0 1 1 1 1 1 \\
0 0 0 0 0 1 0 1 0 0 0 0 0 0 0 0 0 0 0 0 0 1 0 1 0 1 1 0 0 0 1 1 0 0 0 0 \\
0 0 0 0 0 0 1 1 0 0 0 0 0 0 0 0 0 0 0 0 0 0 1 1 0 1 0 1 0 1 1 0 0 0 0 0 \\
0 0 0 0 0 0 0 0 1 0 0 1 0 0 0 0 0 0 0 0 0 0 0 0 0 1 1 0 0 0 1 1 1 0 1 0 \\
0 0 0 0 0 0 0 0 0 1 0 1 0 0 0 0 0 0 0 0 0 0 0 0 0 1 0 1 0 1 1 0 0 0 1 1 \\
0 0 0 0 0 0 0 0 0 0 1 1 0 0 0 0 0 0 0 0 0 0 0 0 0 0 1 1 0 1 0 1 0 1 1 0 \\
0 0 0 0 0 0 0 0 0 0 0 0 1 0 0 1 0 0 0 0 0 1 0 1 0 1 0 1 0 0 0 0 1 0 0 1 \\
0 0 0 0 0 0 0 0 0 0 0 0 0 1 0 1 0 0 0 0 0 0 1 1 0 0 1 1 0 0 0 0 0 1 0 1 \\
0 0 0 0 0 0 0 0 0 0 0 0 0 0 1 1 0 0 0 0 0 1 1 0 0 1 1 0 0 0 0 0 0 0 1 1 \\
0 0 0 0 0 0 0 0 0 0 0 0 0 0 0 0 1 0 0 1 0 1 1 0 0 0 1 1 0 1 1 0 1 0 0 1 \\
0 0 0 0 0 0 0 0 0 0 0 0 0 0 0 0 0 1 0 1 0 1 0 1 0 1 1 0 0 1 0 1 0 1 0 1 \\
0 0 0 0 0 0 0 0 0 0 0 0 0 0 0 0 0 0 1 1 0 0 1 1 0 1 0 1 0 0 1 1 0 0 1 1 \\
0 0 0 0 0 0 0 0 0 0 0 0 0 0 0 0 0 0 0 0 1 1 1 1 0 0 0 0 0 0 0 0 1 1 1 1 \\
0 0 0 0 0 0 0 0 0 0 0 0 0 0 0 0 0 0 0 0 0 0 0 0 1 1 1 1 0 0 0 0 1 1 1 1 \\
0 0 0 0 0 0 0 0 0 0 0 0 0 0 0 0 0 0 0 0 0 0 0 0 0 0 0 0 1 1 1 1 1 1 1 1
\end{array} \right] \]}

{\small \[ G(\mathcal{C}_E^{40}) = \left[ \begin{array} {c}
1 0 0 0 0 0 0 1 0 0 0 1 0 0 0 1 0 0 0 1 0 0 0 1 0 1 1 1 0 0 1 0 0 0 0 1 1 0 1 1 \\
0 1 0 0 0 0 0 1 0 0 0 1 0 0 0 1 0 0 0 1 0 0 0 1 0 0 1 0 0 0 1 0 0 1 0 0 0 0 1 0 \\
0 0 1 0 0 0 0 1 0 0 0 1 0 0 0 1 0 0 0 1 0 0 0 1 0 0 0 1 0 0 1 0 0 1 1 1 0 1 1 1 \\
0 0 0 1 0 0 0 1 0 0 0 1 0 0 0 1 0 0 0 1 0 0 0 1 0 1 0 0 0 0 1 0 0 0 1 0 0 0 0 1 \\
0 0 0 0 1 0 0 1 0 0 0 0 0 0 0 0 0 0 0 0 0 0 0 0 0 1 0 1 0 1 1 0 0 1 1 0 1 0 0 1 \\
0 0 0 0 0 1 0 1 0 0 0 0 0 0 0 0 0 0 0 0 0 0 0 0 0 0 1 1 0 1 0 1 0 1 0 1 0 1 0 1 \\
0 0 0 0 0 0 1 1 0 0 0 0 0 0 0 0 0 0 0 0 0 0 0 0 0 1 1 0 0 0 1 1 0 0 1 1 0 0 1 1 \\
0 0 0 0 0 0 0 0 1 0 0 1 0 0 0 0 0 0 0 0 0 0 0 0 0 1 1 0 0 0 1 1 0 1 0 1 1 1 1 1 \\
0 0 0 0 0 0 0 0 0 1 0 1 0 0 0 0 0 0 0 0 0 0 0 0 0 1 0 1 0 1 1 0 0 0 1 1 0 0 0 0 \\
0 0 0 0 0 0 0 0 0 0 1 1 0 0 0 0 0 0 0 0 0 0 0 0 0 0 1 1 0 1 0 1 0 1 1 0 0 0 0 0 \\
0 0 0 0 0 0 0 0 0 0 0 0 1 0 0 1 0 0 0 0 0 0 0 0 0 0 0 0 0 1 1 0 0 0 1 1 1 0 1 0 \\
0 0 0 0 0 0 0 0 0 0 0 0 0 1 0 1 0 0 0 0 0 0 0 0 0 0 0 0 0 1 0 1 0 1 1 0 0 0 1 1 \\
0 0 0 0 0 0 0 0 0 0 0 0 0 0 1 1 0 0 0 0 0 0 0 0 0 0 0 0 0 0 1 1 0 1 0 1 0 1 1 0 \\
0 0 0 0 0 0 0 0 0 0 0 0 0 0 0 0 1 0 0 1 0 0 0 0 0 1 0 1 0 1 0 1 0 0 0 0 1 0 0 1 \\
0 0 0 0 0 0 0 0 0 0 0 0 0 0 0 0 0 1 0 1 0 0 0 0 0 0 1 1 0 0 1 1 0 0 0 0 0 1 0 1 \\
0 0 0 0 0 0 0 0 0 0 0 0 0 0 0 0 0 0 1 1 0 0 0 0 0 1 1 0 0 1 1 0 0 0 0 0 0 0 1 1 \\
0 0 0 0 0 0 0 0 0 0 0 0 0 0 0 0 0 0 0 0 1 0 0 1 0 1 1 0 0 0 1 1 0 1 1 0 1 0 0 1 \\
0 0 0 0 0 0 0 0 0 0 0 0 0 0 0 0 0 0 0 0 0 1 0 1 0 1 0 1 0 1 1 0 0 1 0 1 0 1 0 1 \\
0 0 0 0 0 0 0 0 0 0 0 0 0 0 0 0 0 0 0 0 0 0 1 1 0 0 1 1 0 1 0 1 0 0 1 1 0 0 1 1 \\
0 0 0 0 0 0 0 0 0 0 0 0 0 0 0 0 0 0 0 0 0 0 0 0 1 1 1 1 0 0 0 0 0 0 0 0 1 1 1 1 \\
0 0 0 0 0 0 0 0 0 0 0 0 0 0 0 0 0 0 0 0 0 0 0 0 0 0 0 0 1 1 1 1 0 0 0 0 1 1 1 1 \\
0 0 0 0 0 0 0 0 0 0 0 0 0 0 0 0 0 0 0 0 0 0 0 0 0 0 0 0 0 0 0 0 1 1 1 1 1 1 1 1
\end{array} \right] \]}

\begin{proposition}
The codes $\mathcal{C}_O^{36}$ and $\mathcal{C}_E^{36}$ are inequivalent binary optimal $[36,19,8]$ linear codes.
\end{proposition}

\noindent\emph{Proof.} Since $C_4^9$ is an additive $(9,2^{10})$ code over $GF(4)$, it follows from Theorem \ref{thm:proj} that $\mathcal{C}_O^{36} = \rho_O(C_4^9)$ and $\mathcal{C}_E^{36} = \rho_E(C_4^9)$ are  binary $[36,19]$ linear codes. It remains to show that the minimum distance is $8$. It is known that codes of these parameters are optimal \cite{Brouwer98}.

Note that any codeword $\mathbf{c} \in \mathcal{C}_O^{36}$ can be written as $\mathbf{c} = \mathbf{a} + \mathbf{b} + \mathbf{d}$ where $\mathbf{a} \in \phi(C_4^9)$, $\mathbf{b}$ is an even sum of weight $4$ vectors whose ones appear in the same column in the projection and $\mathbf{d}$ is equal to either $d_1 = (1000\, \ldots 1000)$ or the zero vector. Since the minimum distance of $C_4^9$ is $4$ and thus $\phi(C_4^9)$ is of minimum distance $8$, $wt(\mathbf{a}) \geq 8$. By the definition of $\mathbf{b}$, it is clear that $wt(\mathbf{b}) \leq 8$. Thus, the minimum distance of $\mathcal{C}_O^{36}$ is at most $8$.

Now partition $\mathbf{c}$ into blocks of length $4$ and call the $ith$ block $\mathbf{c}_i$ (which corresponds to the $ith$ column in the projection). For each $i = 1,\ldots, 9$, write $\mathbf{c}_i = \mathbf{a}_i + \mathbf{b}_i + \mathbf{c}_i$. From the construction, it can be seen that $\mathbf{c}_i = 0000$ if and only if $\mathbf{a}_i = \mathbf{b}_i = \mathbf{d}_i = 0000$.
If $\mathbf{d}$ is the zero vector, i.e., $\mathbf{d}_i = 0000$ for all $i$, then $wt(\mathbf{a}_i+\mathbf{c}_i) \in \{2,4\}$.
Thus, $wt(\mathbf{c}) = wt(\mathbf{a}_i + \mathbf{b}_i) \geq 8$. Suppose $\mathbf{d}_i = 1000$ for all $i$. If $\mathbf{b}_i = 0000$, then $wt(\mathbf{c}_i) = wt(\mathbf{a}_i + \mathbf{d}_i) = 3$. If $\mathbf{b}_i = 1111$, then $wt(\mathbf{a}_i + \mathbf{b}_i + \mathbf{d}_i) \in \{1,3\}$. Since at least $4$ blocks $\mathbf{c}_i$ are nonzero, we conclude that $wt(\mathbf{c}) \geq 8$. Hence the minimum distance of this code is 8.

The case of $\mathcal{C}_E^{36}$ is proved similarly.

Finally, the codes $\mathcal{C}_O^{36}$  and  $\mathcal{C}_E^{36}$ have different weight distributions given in Tables 1 and 2 which was computed by MAGMA.  Therefore they are inequivalent. Furthermore, we have checked by MAGMA that these two codes are not equivalent to the currently known optimal $[36, 19, 8]$ code denoted by $BKLC(GF(2), 36, 19)$ in the MAGMA database because $BKLC(GF(2), 36, 19)$ has $A_8=1033$. Note that the codes $\mathcal{C}_O^{36}$  and  $\mathcal{C}_E^{36}$ have the automorphism groups of order 6, while $BKLC(GF(2), 36, 19)$ has a trivial automorphism.
\hfill $\blacksquare$

\bigskip

For length 40, we have a similar result as in Proposition 2.
We display the weight distributions of $C_O^{40}$ and $C_E^{40}$ in Tables 3 and 4. Furthermore, we have checked by MAGMA that these two codes are not equivalent to the currently known optimal $[40, 22, 8]$ code denoted by $BKLC(GF(2), 40, 22)$ in the MAGMA database because $BKLC(GF(2), 40, 22)$ has $A_8=1412$.

\begin{proposition}
The codes $\mathcal{C}_O^{40}$ and $\mathcal{C}_E^{40}$ are inequivalent binary optimal $[40,22,8]$ linear codes.
\end{proposition}

\begin{table}[H]
\centering {\small
\begin{tabular}{|l|l|l|l|l|}
\hline
$A_0=1$ & $A_8=444$ & $A_9=496$ & $A_{10}=2160$ & $A_{11}=4752$ \\
$A_{12}=8760$ & $A_{13}=17856$ & $A_{14}=28992$ & $A_{15}=44352$ & $A_{16}=54318$ \\
$A_{17}=62496$ & $A_{18}=72864$ & $A_{19}=66528$ &
$A_{20}=54192$ & $A_{21}=41664$ \\
$A_{22}=28992$ & $A_{23}=19008$ & $A_{24}=8844$ & $A_{25}=4464$ &
$A_{26}=2160$ \\
 $A_{27}=528$ & $A_{28}=408$ & $A_{32}=9$ & & \\
\hline
\end{tabular} } \caption{Weight distribution of $C_O^{36}$} \label{tab:wd-36-O}
\end{table}

\begin{table}[H]
\centering {\small
\begin{tabular}{|l|l|l|l|l|}
\hline
$A_0=1$ &  $A_8=444$ & $A_9=528$ & $A_{10}=2160$ & $A_{11}=4464$ \\
$A_{12}=8760$ & $A_{13}=19008$ &
$A_{14}=28992$ & $A_{15}=41664$ & $A_{16}=54318$ \\
$A_{17}=66528$ & $A_{18}=72864$ & $A_{19}=62496$ &
$A_{20}=54192$ & $A_{21}=44352$ \\
 $A_{22}=28992$ & $A_{23}=17856$ & $A_{24}=8844$ & $A_{25}=4752$ &
$A_{26}=2160$ \\
 $A_{27}=496$ & $A_{28}=408$ & $A_{32}=9$ & &\\
\hline
\end{tabular} } \caption{Weight distribution of $C_E^{36}$} \label{tab:wd-36-E}
\end{table}

\begin{table}[H]
\centering {\small
\begin{tabular}{|l|l|l|l|l|}
\hline
$A_0=1$ & $A_8=741$ & $A_{10}=6144$ & $A_{12}=42736$ & $A_{14}=176640$ \\
 $A_{16}=484890$  & $A_{18}=849408$ &  $A_{20}=1073184$ & $A_{22}=849408$ & $A_{24}=484890$ \\
$A_{26}=176640$ & $A_{28}= 42736$ &  $A_{30}=6144$ & $A_{32}=741$  & $A_{40}=1$\\
\hline
\end{tabular} } \caption{Weight distribution of $C_O^{40}$} \label{tab:wd-40-O}
\end{table}

\begin{table}[H]
\centering {\small
\begin{tabular}{|l|l|l|l|l|}
\hline
$A_0=1$ &   $A_8=741$ &  $A_{10}=6208$ &  $A_{12}=42096$  &  $A_{14}=179520$ \\
$A_{16}=477210$ &  $A_{18}=
862848$ & $A_{20}=1057056$ &  $A_{22}=862848$  &  $A_{24}=477210$ \\
 $A_{26}=179520$ & $A_{28}=42096$ &
$A_{30}=6208$ &  $A_{32}=741$  &  $A_{40}=1$ \\
\hline
\end{tabular} } \caption{Weight distribution of $C_E^{40}$} \label{tab:wd-40-E}
\end{table}

\section{Projection decoding}\label{sec:decoding}

Let $v \in GF(2)^{4m}$ and $\mathbf{v}$ its associated array, defined in Section \ref{sec:projection}.
From this, we can partition the elements of $GF(2)^4$ with respect to its inner product with the row labels, as follows:
\begin{table}[H]
\centering {\small
\begin{tabular}{|c|c|}
\hline
$0 \mapsfrom \begin{array}{c} 0 \\ 0 \\ 0 \\ 0\end{array}, \begin{array}{c} 1 \\ 1 \\ 1 \\ 1 \end{array},  \begin{array}{c}  1 \\ 0 \\ 0 \\ 0 \end{array}, \begin{array}{c} 0 \\ 1 \\ 1 \\ 1 \end{array}$
 & $1 \mapsfrom \begin{array}{c} 1 \\ 1 \\0 \\ 0\end{array}, \begin{array}{c} 0 \\ 0 \\ 1 \\ 1 \end{array}, \begin{array}{c} 0 \\ 1 \\ 0 \\ 0 \end{array}, \begin{array}{c} 1 \\ 0 \\ 1 \\ 1 \end{array}$  \tabularnewline
\hline
$\omega \mapsfrom \begin{array}{c} 1 \\ 0 \\ 1 \\ 0\end{array}, \begin{array}{c} 0 \\ 1 \\ 0 \\ 1 \end{array},  \begin{array}{c}  0 \\ 0 \\ 1 \\ 0 \end{array}, \begin{array}{c} 1 \\ 1 \\ 0 \\ 1 \end{array}$
 & $\overline{\omega} \mapsfrom \begin{array}{c} 1 \\ 0 \\0 \\ 1\end{array}, \begin{array}{c} 0 \\ 1 \\ 1 \\ 0 \end{array}, \begin{array}{c} 0 \\ 0 \\ 0 \\ 1 \end{array}, \begin{array}{c} 1 \\ 1 \\ 1 \\ 0 \end{array}$ \tabularnewline
\hline
\end{tabular} } \caption{Projection of $GF(2)^4$ onto $GF(4)$} \label{tab:F2F4}
\end{table}

From Definitions \ref{def:projectionO} and \ref{def:projectionE}, we know that if $v\in\mathcal{C}$, where $\mathcal{C}$
is a code with projection $O$ on to $C_{4}$, then the columns of $\mathbf{v}$,
as well as the first row have the same parity.
Before we give the decoding algorithm, we first take note of the following observations regarding the error positions in the array $\mathbf{v}$.

\medskip
\noindent{\bf Remarks} \begin{enumerate}
\item An error on the first row of $\mathbf{v}$ preserves $\Proj(\mathbf{v})$.
\item An error on the coordinate that is not in the first row definitely changes $\Proj(\mathbf{v})$.
\item Two errors in the same column definitely changes $\Proj(\mathbf{v})$.
\item Three errors in the same column preserves $\Proj(\mathbf{v})$ if and only if the first entry is not changed.
\end{enumerate}

We now present a simple decoding algorithm, which we will call the \emph{projection decoding}, that can be used for any binary linear code with projection $O$ or projection $E$ onto some additive code $C_4$ over $GF(4)$.  This decoding algorithm can correct errors of weight up to three. The idea is to use syndrome decoding in $C_4$, which has shorter length to locate errors in the binary codeword. The decoding is then completed by changing the columns in the array corresponding to the corrected entry in the projection, by using Table \ref{tab:F2F4}.

Let $\mathcal{C}$ be a binary linear code of length $4n$ with projection $O$ or projection $E$ onto $C_4$, an additive code over $GF(4)$. Let $y$ be the received vector and assume that $y$ is a noisy codeword of $\mathcal{C}$
with at most $3$ errors. Let $\mathbf{y}$ be the associated array and  $\overline{\mathbf{y}} = \Proj(\mathbf{y})$. Denote by $\mathbf{y}_i$ the $i$th column of $\mathbf{y}$ and by $\overline{\mathbf{y}}_i$ the $i$th entry of $\overline{\mathbf{y}}$. Let $H$ be the parity-check matrix of $C_{4}$
and denote the $i$th column of $H$ by $H_{i}$. The projection decoding of $\mathcal{C}$ is carried out as follows.

\bigskip
\noindent{\bf Projection decoding algorithm}

\begin{enumerate}
\item Check the parity of the columns and the first row of $\mathbf{y}$.
\item Let $y^o$, $y^e$ be the number of columns of $\mathbf{y}$ with odd or even parity,
resp., and let $p=\min(y^o,y^e)$.
\item Depending on $p$, perform the following:
	\begin{enumerate}
	\item If $p=0$, compute the syndrome $\mathbf{s} = \overline{\mathbf{y}}H^T$.
		\begin{enumerate}
		\item If $\mathbf{s}=\mathbf{0}$, then we say that no error occurred.
		\item If $\mathbf{s}\neq \mathbf{0}$, the syndrome is a scalar multiple of one of
		the columns of $H$, say $\mathbf{s}=e_{i}H_{i}$ for some $e_{i}\in GF(4)$.
		Hence, the two errors occurred on the $i$th column of $\mathbf{y}$.
		Replace the $i$th coordinate of $\overline{\mathbf{y}}$ by $\overline{\mathbf{y}_{i}}+e_{i}\in GF(4)$
		and replace the $i$th column of $\mathbf{y}$ with the vector corresponding
		to $\overline{\mathbf{y}_{i}}+e_{i}$ (see Table \ref{tab:F2F4}) such that the parity conditions are
		satisfied.
		\end{enumerate}
	\item If $p=1$, let $\mathbf{y}_{i}$ be the column with the different parity. Compute
	the syndrome $\mathbf{s} = \overline{\mathbf{y}}H^T$.
		\begin{enumerate}
		\item If $\mathbf{s}=\mathbf{0}$, check the parity of the first row.
			\begin{enumerate}
			\item If the parity of the first row is the same as the parity of $\mathbf{y_i}$, then one error occurred on 			the first entry of $\mathbf{y}_i$. Change this entry to decode $y$.
			\item If the parity of the first row is different from the parity of $\mathbf{y_i}$, then three errors occurred 			on the 2nd to 4th entries of $\mathbf{y_i}$. Change these entries to decode $y$.
			\end{enumerate}
		\item If $\mathbf{0} \neq \mathbf{s} = e_{i}H_{i}$, then one or three errors occurred in the
		$i$th column of $\mathbf{y}$. Replace the $i$th coordinate of $\overline{\mathbf{y}}$
		by $\overline{\mathbf{y}_{i}}+e_{i}\in GF(4)$ and replace the $i$th column
		of $\mathbf{y}$ with the vector corresponding to $\overline{\mathbf{y}_{i}}+e_{i}$
		such that the parity conditions are satisfied.
		\item If $\mathbf{s}=e_{j}H_{j}$ for some $j\neq i$, then two errors occurred
		in column $j$ and one error in the first entry of column $i$. Replace the $jth$ coordinate
		of $\overline{\mathbf{y}}$ by $\overline{\mathbf{y}_{j}}+e_{j}\in GF(4)$ and
		replace the $jth$ column of $\mathbf{y}$ with the vector corresponding
		to $\overline{\mathbf{y}_{j}}+e_{j}$ having the same parity as $\mathbf{y}_{j}$
		and of distance $2$ from $\mathbf{y}_{j}$. Finally, replace the first entry of $\mathbf{y_{i}}$
		and check that
		the parity conditions are satisfied.
		\end{enumerate}
	\item If $p=2$, let $\mathbf{y}_{i},\mathbf{y}_{j}$ be the columns with the different parity.
	Compute the syndrome $\mathbf{s} = \overline{\mathbf{y}}H^T$. We know that $\mathbf{s}=e_{i}H_{i}+e_{j}H_{j}$.
		\begin{enumerate}
		\item If $\mathbf{s}=0$, then the two errors occurred on the first coordinates
		of $\mathbf{y}_{i}$ and $\mathbf{y}_{j}$. Replace both coordinates
		and check if the parity conditions are satisfied.
		\item If $e_{i}\neq0$ and $e_{j}=0$, then error occurred in the $i$th
		column of $\mathbf{y}$ and the first coordinate of $y_{j}$. Replace
		the first coordinate of $y_{j}$. Then replace the $i$th coordinate
		of $\overline{\mathbf{y}}$ by $\overline{\mathbf{y}_{i}}+e_{i}\in GF(4)$ and
		replace the $i$th  of $\mathbf{y}$ with the vector corresponding
		to $\overline{\mathbf{y}_{i}}+e_{i}$ such that the parity conditions are
		satisfied.
		\item If $e_{i},e_{j}\neq0$, then error occurred in the $i$thand $j$th
		column of $\mathbf{y}$. Replace the $i$th coordinate of $\overline{\mathbf{y}}$
		by $\overline{\mathbf{y}_{i}}+e_{i}\in GF(4)$ and the $j$th coordinate
		by $\overline{\mathbf{y}_{j}}+e_{j}\in GF(4)$ then replace the $i$th
		and $j$th columns of $\mathbf{y}$ with the corresponding vectors
		such that the parity conditions are satisfied.
		\end{enumerate}
	\item If $p=3$, let $\mathbf{y}_{i},\mathbf{y}_{j}$ and $\mathbf{y}_{k}$ be the columns with different
	parity. Then there are one error each on these columns. Compute the
	syndrome $\mathbf{s} = \overline{\mathbf{y}}H^T$. We know that $\mathbf{s}=e_{i}H_{i}+e_{j}H_{j}+e_{k}H_{k}$
		\begin{enumerate}
		\item If $\mathbf{s} = \mathbf{0} $, then the three errors occurred on the first coordinates
		of $\mathbf{y}_{i}, \mathbf{y}_{j}$ and $\mathbf{y}_{k}$. Replace
		the first coordinates in these columns and check if the parity conditions
		are satisfied.
		\item If $e_{i}\neq0$ and $e_{j},e_{k} = 0$, then one error occurred in
		the $i$thcolumn of $\mathbf{y}$ and and one error each on the the first coordinates of $\mathbf{y}_{j}$
		and $\mathbf{y}_{k}$. Replace the first coordinates of $\mathbf{y}_{j}$
		and $\mathbf{y}_{k}$. Then replace the $i$th coordinate of $\overline{\mathbf{y}}$
		by $\overline{\mathbf{y}_{i}}+e_{i}\in GF(4)$ and replace the $i$th column
		of $\mathbf{y}$ with the vector corresponding to $\overline{\mathbf{y}_{i}}+e_{i}$
		such that the parity conditions are satisfied.
		\item If $e_{i}\neq0,e_{j}\neq0$ and $e_{k}=0$, then one error each occurred
		in the $i$th and $j$th column of $\mathbf{y}$ and the first coordinate
		of $\mathbf{y}_{k}$. Replace the first coordinate of $\mathbf{y}_{k}$.
		Then replace the $i$th coordinate of $\overline{\mathbf{y}}$ by $\overline{\mathbf{y}_{i}}+e_{i}\in GF(4)$
		and the $j$th coordinate by $\overline{\mathbf{y}_{j}}+e_{j}\in GF(4)$
		then replace the $i$th and $j$th columns of $\mathbf{y}$ with the
		corresponding vectors such that the parity conditions are satisfied.
		\item If $e_{i},e_{j},e_{k}\neq0$, then let $\mathbf{y}'=\overline{\mathbf{y}}+\mathbf{e}$
		where $\mathbf{e}$ is the vector of length $9$ with $e_{i},e_{j}$
		and $e_{k}$ on the $i,j$ and $k$th coordinate and zero elsewhere.
		Replace one coordinate each in the $i,j,k$ columns of $\mathbf{y}$
		so that it becomes $\mathbf{y}'$ and the parity conditions are satisfied.
		\end{enumerate}
	\end{enumerate}
\end{enumerate}

\medskip
\noindent{\bf Remarks}
We remark that our algorithm for codes $\mathcal{C}_O^{36}$,  $\mathcal{C}_E^{36}$, $\mathcal{C}_O^{40}$, and $\mathcal{O}_E^{40}$ is complete and ending because we have considered all possible number of column parities $p=\min(y^o,y^e) \le 3$ and adjusted at most three errors according to the top row parity so that Propositions 1 and 2 are satisfied.

\section{Examples}
In this section, we provide examples to illustrate how the given decoding algorithm works. Even though these are samples, most remaining cases are done similarly.

\medskip

The following two examples illustrate how to decode $\mathcal{C}_O^{36}$ by hand. As a linear code over $GF(4)$, $C_4^9$ has the following parity check matrix.
\[ H(C_4^9) = \left[ \begin{array}{ccccccccc}
1 & 0 & 0 & 0 & 1 & 1 & \overline{\omega} & 0 & 1 \\
0 & 1 & 0 & 0 & 1 & 0 & \omega & \overline{\omega} &  1 \\
0 & 0 & 1 & 0 & \omega & \overline{\omega} &  1 & \omega & 1 \\
0 & 0 & 0 & 1 & 1 & \overline{\omega} & 0 & 1 & \overline{\omega}
\end{array} \right] \]
as a parity check matrix for $C_4^9$.

\bigskip
\noindent{\bf Example 1} In this example, we illustrate the projection decoding of $\mathcal{C}_O^{36}$. Let $y \in GF(2)^{36}$ be a codeword in $\mathcal{C}_O^{36}$ with some error of weight up to three. Consider the following projection of $y$.
\[\mathbf{y}=\begin{array}{c|ccccccccc}
 & 1 & 2 & 3 & 4 & 5 & 6 & 7 & 8 & 9\\
\hline
0 & 0 & 1 & 0 & 1 & 0 & 1 & 0 & 1 & 0\\
1 & 0 & 1 & 1 & 0 & 1 & 0 & 0 & 1 & 1\\
\omega & 1 & 1 & 0 & 0 & 0 & 1 & 1 & 0 & 0\\
\overline{\omega}  & 0 & 0 & 0 & 0 & 1 & 0 & 0 & 1 & 0  \\ \hline
\overline{\mathbf{y}} =   & \omega & \overline{\omega} & 1 & 0 & \omega & \omega & \omega & \omega & 1
\end{array}\]
All the columns have odd parity except for the 5th and 6th columns, and hence $p=2$. Then we have the syndrome \[\mathbf{s} = \left[ \begin{array}{c} \omega \\ \omega \\ \overline{\omega} \\ \omega \end{array} \right] = \omega H(C_4^9)_5. \]
Therefore we proceed to ii of Step 3(c) with $e_5 = \omega$ and $e_6 = 0$. We replace $\overline{\mathbf{y}}_5 = \omega$ by $ \overline{\mathbf{y}}_5 + e_5 = \omega + \omega = 0$ and then replace $\mathbf{y}_5 = {\small \left[ \begin{array}{c} 0 \\ 1 \\ 0 \\ 1  \end{array} \right] }$ by a vector in Table \ref{tab:F2F4} corresponding to $0$ closest to it. Finally, we replace the first entry of $\mathbf{y}_6$. So there are two errors in $y$. Therefore, we have decoded $y$ as
\[\mathbf{y}=\begin{array}{c|ccccccccc}
 & 1 & 2 & 3 & 4 & 5 & 6 & 7 & 8 & 9\\
 \hline
0 & 0 & 1 & 0 & 1 & \mathbf{0} & \mathbf{0} & 0 & 1 & 0\\
1 & 0 & 1 & 1 & 0 & \mathbf{1} & 0 & 0 & 1 & 1\\
\omega & 1 & 1 & 0 & 0 & \mathbf{1} & 1 & 1 & 0 & 0\\
\overline{\omega}  & 0 & 0 & 0 & 0 & \mathbf{1} & 0 & 0 & 1 & 0  \\ \hline
  & \omega & \overline{\omega} & 1 & 0 & \mathbf{0} & \omega & \omega & \omega & 1
\end{array}\]

\bigskip
\noindent{\bf Example 2} Let $y \in GF(2)^{36}$ be a codeword of $\mathcal{C}_E^{36}$ with some error of weight up to three and have the following projection
\[
\mathbf{y}=\begin{array}{c|ccccccccc}
 & 1 & 2 & 3 & 4 & 5 & 6 & 7 & 8 & 9\\
 \hline
0 & 0 & 1 & 0 & 0 & 0 & 1 & 0 & 1 & 0\\
1 & 0 & 1 & 1 & 0 & 0 & 0 & 0 & 0 & 1\\
\omega & 1 & 0 & 1 & 0 & 0 & 1 & 0 & 0 & 0\\
\overline{\omega}  & 0 & 1 & 1 & 1 & 1 & 1 & 1 & 1 & 0  \\ \hline
\overline{\mathbf{y}} =   & \omega & \omega & 0 & \overline{\omega} & \overline{\omega} & 1 & \overline{\omega} & \overline{\omega} & 1
\end{array}
\]
Note that $p=1$ and the parity of the 8th column differs from the rest of the column, i.e. $i=8$ in Step 3(b). We then compute the syndrome:
\[\mathbf{s} = \left[\begin{array}{c} \overline{\omega} \\ \overline{\omega} \\ 1 \\ \overline{\omega} \end{array} \right] = \overline{\omega} \left[ \begin{array}{c} 1 \\ 1 \\ \omega \\ 1 \end{array} \right] = \overline{\omega} H(C_4^9)_5.\]
So we have $j=5\neq i$ and $e_5 = \overline{\omega}$. Proceeding with iii of Step 3(b), we replace $\overline{\mathbf{y}}_5 = \overline{\omega}$ by $\overline{\mathbf{y}}_5 + e_5 = \overline{\omega} + \overline{\omega} = 0$ and replace the fifth column $\mathbf{y}_5 = {\small \left[ \begin{array}{c} 0 \\ 0 \\ 0 \\ 1 \end{array} \right] }$ by a vector in Table \ref{tab:F2F4} corresponding to 0 which is of the same parity and distance 2 from $\mathbf{y}_5$. Then we change the first entry of $\mathbf{y}_8$. There are three errors in $y$. We have decoded $y$ as
\[
\mathbf{y}=\begin{array}{c|ccccccccc}
 & 1 & 2 & 3 & 4 & 5 & 6 & 7 & 8 & 9\\
 \hline
0 & 0 & 1 & 0 & 0 & \mathbf{0} & 1 & 0 & \mathbf{0} & 0\\
1 & 0 & 1 & 1 & 0 & \mathbf{1} & 0 & 0 & 0 & 1\\
\omega & 1 & 0 & 1 & 0 & \mathbf{1} & 1 & 0 & 0 & 0\\
\overline{\omega}  & 0 & 1 & 1 & 1 & \mathbf{1} & 1 & 1 & 1 & 0  \\ \hline
  & \omega & \omega & 0 & \overline{\omega} & \mathbf{0} & 1 & \overline{\omega} & \overline{\omega} & 1
\end{array}
\]

\bigskip
Next we show examples of projection decoding of $\mathcal{C}_O^{40}$ and $\mathcal{C}_E^{40}$. We use the following parity check for $C_4^{10}$.
\[ H(C_4^{10}) = \left[ \begin{array}{cccccccccc}
 1 &  0 &  0 &  0 &  1 &  1 & \overline{\omega} &  0 &  1 & \overline{\omega} \\
 0 &  1 &  0 &  0 &  1 &  0 &  \omega & \overline{\omega} & 1 &  \omega \\
 0 &  0 &  1 &  0 &  \omega & \overline{\omega} &  1 &  \omega &  1 &  0 \\
 0 &  0 &  0 &  1 &  1 & \overline{\omega} &   0 &  1 & \overline{\omega} &  w
 \end{array} \right] \]

\noindent{\bf Example 3} Assume $y \in GF(2)^{40}$ is a codeword of $\mathcal{C}_O^{40}$ with error of weight up to three and having the following projection
\[
\mathbf{y}=\begin{array}{c|cccccccccc}
 & 1 & 2 & 3 & 4 & 5 & 6 & 7 & 8 & 9 & 10 \\
 \hline
0 & 1 & 0 & 1 & 0 & 0 & 0& 0 & 1 & 1 & 1 \\
1 & 0 & 0 & 1 & 0 & 1 & 1 & 1 & 0 & 0 & 1\\
\omega & 0 & 0 & 1 & 0 & 1 & 0 & 1 & 0 & 0 & 0 \\
\overline{\omega}  & 0 & 1 & 0 & 1 & 1 & 0 & 1 & 0 & 0 & 1  \\ \hline
\overline{\mathbf{y}} =   & 0 & \overline{\omega} & \overline{\omega} & \overline{\omega} & 0 & 1 & 0 & 0 & 0 &  \omega
\end{array} \]
All the columns and the first row are of odd parity, so $p=0$. Computing the syndrome, we have \[ \mathbf{s} =  \left[ \begin{array}{c} 0 \\ 0 \\ 0 \\ \overline{\omega} \end{array} \right] = \overline{\omega} \left[ \begin{array}{c} 0 \\ 0 \\ 0 \\ 1 \end{array} \right] = \overline{\omega} H(C_4^{10})_4. \]
Since $\mathbf{s} \neq \mathbf{0}$, by ii of Step 3(a) with $e_4 = \overline{\omega}$, two errors occurred on the 4th column of $\mathbf{y}$. We replace $\overline{\mathbf{y}}_4 = \overline{\omega}$ by $\overline{\mathbf{y}}_4 + e_4 = \overline{\omega} + \overline{\omega} = 0$ and replace the fourth column $\mathbf{y}_4$ by the vector in Table \ref{tab:F2F4} corresponding to $0$ and of distance $2$ from $\mathbf{y}_4 = {\small \left[ \begin{array}{c} 0 \\ 0 \\ 0 \\ 1 \end{array} \right] }$. There are two errors in $y$. Therefore, $y$ is decoded as
\[
\mathbf{y'}=\begin{array}{c|cccccccccc}
 & 1 & 2 & 3 & 4 & 5 & 6 & 7 & 8 & 9 & 10 \\
 \hline
0 & 1 & 0 & 1 & \mathbf{0} & 0 & 0& 0 & 1 & 1 & 1 \\
1 & 0 & 0 & 1 & \mathbf{1} & 1 & 1 & 1 & 0 & 0 & 1\\
\omega & 0 & 0 & 1 & \mathbf{1} & 1 & 0 & 1 & 0 & 0 & 0 \\
\overline{\omega}  & 0 & 1 & 0 & \mathbf{1} & 1 & 0 & 1 & 0 & 0 & 1  \\ \hline
  & 0 & \overline{\omega} & \overline{\omega} & \mathbf{0} & 0 & 1 & 0 & 0 & 0 &  \omega
\end{array}. \]

\bigskip
\noindent{\bf Example 4} Assume $y \in GF(2)^{40}$ is a codeword of $\mathcal{C}_E^{40}$ with some error of weight up to $3$. Let $y$ have the following projection
\[
\mathbf{y}=\begin{array}{c|cccccccccc}
 & 1 & 2 & 3 & 4 & 5 & 6 & 7 & 8 & 9 & 10 \\
 \hline
0 & 0 & 0 & 1 & 0 & 1 & 0 & 0 & 1 & 0 & 0 \\
1 & 1 & 0 & 0 & 0 & 0 & 0 & 1 & 1 & 1 & 0 \\
\omega & 0 & 1 & 1 & 1 & 0 & 1 & 0 & 1 & 0 & 1 \\
\overline{\omega} & 1 & 1 & 1 & 1 & 1 & 1 & 1 & 0 & 1 & 0 \\ \hline
\overline{\mathbf{y}} = & \omega & 1 & 1 & 1 & \overline{\omega} & 1 & \omega & \overline{\omega} & \omega & \omega
\end{array} \]
All the columns except columns 3, 8 and 10 are even so $p=3$. The syndrome is
\[ \mathbf{s} = \left[ \begin{array}{c} \omega \\ \omega \\ \omega \\ 0 \end{array} \right] =  H(C_4^{10})_8 + \overline{\omega} H(C_4^{10)})_{10} \]
Hence, from iii of Step 3(d), we have $e_8 = 1$, $e_{10} = \overline{\omega}$ and $e_3 = 0$. So we replace $\overline{\mathbf{y}}_8$ by $\overline{\mathbf{y}}_8 + e_8 = \overline{\omega} + 1 = \omega$ and $\mathbf{y}_8$ by the vector corresponding to $\omega$ in Table \ref{tab:F2F4} closest to {\small $\left[ \begin{array}{c} 1 \\ 1 \\ 1 \\ 0 \end{array} \right]$}. We also replace $\overline{\mathbf{y}}_{10}$ by $\overline{\mathbf{y}}_{10} + e_{10} = \omega + \overline{\omega} = 1$ and $\mathbf{y}_{10} = {\small \left[ \begin{array}{c} 0 \\ 0 \\ 1 \\ 0 \end{array} \right]} $ by the vector closest to it corresponding to $1$ in Table \ref{tab:F2F4}. Finally we replace the first entry of column 3. There are three errors in $y$. Therefore, $y$ is decoded as
\[
\mathbf{y}=\begin{array}{c|cccccccccc}
 & 1 & 2 & 3 & 4 & 5 & 6 & 7 & 8 & 9 & 10 \\
 \hline
0 & 0 & 0 & \mathbf{0} & 0 & 1 & 0 & 0 & \mathbf{1} & 0 & \mathbf{0} \\
1 & 1 & 0 & 0 & 0 & 0 & 0 & 1 & \mathbf{0} & 1 & \mathbf{0} \\
\omega & 0 & 1 & 1 & 1 & 0 & 1 & 0 & \mathbf{1} & 0 & \mathbf{1} \\
\overline{\omega} & 1 & 1 & 1 & 1 & 1 & 1 & 1 & \mathbf{0} & 1 & \mathbf{1} \\ \hline
 & \omega & 1 & 1 & 1 & \overline{\omega} & 1 & \omega & \pmb{\omega} & \omega & \mathbf{1}
\end{array} \]

\noindent{\bf{Remark}}
The most time consuming part (or the dominating complexity part of the algorithm) is to decode the projected vector $\overline{\mathbf{y}}$ in the linear code $\mathcal{C}_4^9$ or $\mathcal{C}_4^{10}$ using the syndrome decoding. However, since we know which positions (or columns) have errors, this syndrome decoding can be done in at most $3 \times 3 \times 3=27$ possible ways because there are scalar multiples in the syndrome equation involving three nonzero scalars in Step (d) of the algorithm.

\medskip

\section{Conclusion}
In this paper, we have described how to decode binary optimal $[36,19,8]$ and $[40,22,8]$ codes by projecting them onto the linear $[9,5,4]$ and $[10, 6, 4]$ codes over $GF(4)$. Even though there were similar decoding for self-dual codes of lengths 24, 32, 40, there was no known efficient decoding algorithm for these non-self-dual optimal codes. Actually our algorithm works for any linear code with a projection onto a linear or additive code over $GF(4)$ for errors of weight at most $3$.

\end{document}